\documentclass[%
 reprint,
 amsmath,amssymb,
 aps,
]{revtex4-2}

\usepackage{graphicx}
\usepackage{dcolumn}
\usepackage{bm}
\usepackage{amsmath,amssymb}
\usepackage{color} 
\usepackage{booktabs} 

\begin{document}

\preprint{APS/123-QED}

\title{Two-dimensional hydrodynamic simulation for synchronization in coupled density oscillators}

\author{Nana Takeda}
\author{Hiroaki Ito}
\author{Hiroyuki Kitahata}
\email{kitahata@chiba-u.jp}
\affiliation{
 Department of Physics, Chiba University, Chiba 263-8522, Japan
}

\date{\today}

\begin{abstract}
A density oscillator is a fluid system in which oscillatory flow occurs between different density fluids through the pore connecting them.
We investigate the synchronization in coupled density oscillators using two-dimensional hydrodynamic simulation and analyze the stability of the synchronous state based on the phase reduction theory.
Our results show that the antiphase, three-phase, and 2-2 partial-in-phase synchronization modes spontaneously appear as stable states in two, three, and four coupled oscillators, respectively.
The phase dynamics of coupled density oscillators is interpreted with their sufficiently large first Fourier components of the phase coupling function.
\end{abstract}

\maketitle


\section{introduction}

Synchronization of coupled limit-cycle oscillators is observed in various systems such as electrical, chemical, and biological systems \cite{Winfree,Kuramoto,Strogatz}.
Synchronization among many oscillators makes the oscillation more robust in some systems, for example, an ensemble of pacemaker cells in a heart beating, whereas it can be harmful in other systems, for example, an ensemble of neurons in an epileptic seizure \cite{Strogatz}.
Therefore, it is important to understand the mechanism of synchronization and to control the synchronous behavior as needed.
Many efforts have been made to describe synchronization theoretically.
One of the well-known methods is a phase reduction theory, which approximately describes the behavior of weakly coupled oscillators using one variable (i.e., phase) for each oscillator \cite{Kuramoto}.
The phase coupling function characterizing the interaction in coupled oscillators has been investigated for various systems \cite{Stankovski}.
For fluid systems, for example, the oscillatory convection in a Hele-Shaw cell, flow around the beating flagella, and the cylinder wake were studied, and their synchronization under hydrodynamic interaction was discussed with phase reduction \cite{Kawamura1,Kawamura2,Taira}.
The relation between synchronization and energy efficiency in fluid systems was also studied to understand how the emerging synchronization mode is selected as an optimal state \cite{Attard,cilia1,cilia2}.
The energy efficiency can be evaluated from some criteria with energy dissipation, such as viscous dissipation and the amount of fluid transport.
The investigations of the phase dynamics and energy efficiency will be effective for understanding the underlying mechanism of synchronization phenomena in fluid systems.

A density oscillator is a typical fluid system that exhibits limit-cycle oscillations.
It consists of the lower density fluid in the outer container and the higher density fluid in the inner container with a pore at the bottom.
In appropriate conditions for the fluid densities and the dimensions of the containers, the downward flow of the higher density fluid and the upward flow of the lower density fluid through the pore alternate periodically.
This oscillatory phenomenon was first reported by Martin in 1970 \cite{Martin}.
Subsequently, experimental and theoretical studies on a density oscillator were reported from various aspects, such as the mechanism of the oscillatory flows \cite{Alf,Noyes,Upa,Steinbock,Okamura,Aoki,Yoshikawa,Kano1,Kano2,Kenfack1}, the bifurcation phenomena \cite{Aoki,Kenfack2,Ito,Takeda}, and the response to external force \cite{Gonza,Kenfack3}.
In coupled density oscillators, which consist of multiple inner containers in a common outer container, various synchronization modes are observed depending on the coupling strength and detuning of the intrinsic frequencies \cite{Miyakawa1,Nakata,Yoshikawa,Kano1,Kano2,Miyakawa2,Horie,Kenfack1}.
For almost equal intrinsic frequencies, the two coupled oscillators exhibit the antiphase synchronization mode \cite{Nakata,Yoshikawa,Kano2,Horie}, the three coupled ones exhibit the three-phase synchronization mode (also called the rotation mode) \cite{Yoshikawa,Kano2,Miyakawa2,Horie}, and the four coupled ones exhibit several synchronization modes depending on the coupling strength \cite{Miyakawa2,Horie} in experiments.
Several theoretical models for the oscillatory flows reproduced the synchronization modes \cite{Yoshikawa,Kano1,Kano2,Kenfack1,Horie}.
Horie \textit{et al.} represented the time variation of the water level using a combination of exponential functions and investigated synchronization modes in two, three, and four coupled oscillators with the phase model \cite{Horie}.
They mainly focused on clustering states depending on the parameters, and did not perform detailed analysis of the stability for all the possible synchronization modes.
In order to understand the criteria for selection of the emerging synchronization mode, we need to compare hydrodynamic behaviors between stable and unstable synchronization modes, as well as to analyze the stability of them.

In the present study, we investigate the hydrodynamic behavior and phase dynamics of coupled density oscillators.
We take the following three approaches.
First, we perform the two-dimensional hydrodynamic simulation for coupled density oscillators consisting of two, three, and four inner containers and observe the synchronization phenomena.
The time evolutions of the water levels and the phase differences are calculated for stable and unstable synchronization modes.
Second, the phase response to the perturbation is measured by the hydrodynamic simulation for a single density oscillator.
Based on the phase reduction theory, the dynamics of the phase differences between coupled oscillators is analyzed.
Third, a linear stability analysis is applied for the fixed points of the phase differences corresponding to stable and unstable synchronization modes.
We evaluate the stabilities of the fixed points generically with the Fourier components of the phase coupling function.
We finally discuss the criterion for selection of the synchronization mode.

\section{simulation model}

\begin{figure}[tb]
  \begin{center}
    \includegraphics{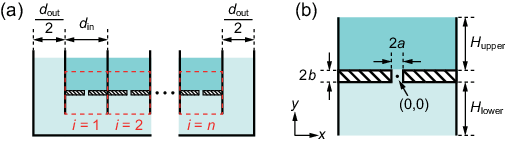}
  \end{center}
  \caption{Two-dimensional model of coupled density oscillators. (a) Schematic drawing of coupled density oscillators consisting of $n$ inner containers. The red broken rectangles represent the calculation areas. (b) Details of the calculation area. The hatched area shows the bottom wall of the inner container.}
  \label{fig:system}
\end{figure}

We perform the two-dimensional hydrodynamic simulation for coupled density oscillators consisting of $n$ identical inner containers ($n=1,2,3,4$).
The simulation model is extended from the model for a single density oscillator introduced in our previous study \cite{Takeda}.
Figure~\ref{fig:system}(a) shows the schematic drawing of coupled density oscillators.
The widths of the inner and outer containers, $d_\mathrm{in}$ and $d_\mathrm{out}$, are fixed.
The calculation areas are fixed inside the fluid and separated by the walls for each inner container.
The calculations are performed in respective areas.
The $n$ oscillators interact only through the common pressure at the lower boundaries of the calculation areas; this pressure corresponds to the water level in the common outer container.
Figure~\ref{fig:system}(b) shows the $i$th calculation area ($i=1,2,\ldots,n$).
The width and length of the pore through which the fluid passes are $2a$ and $2b$, respectively.
The distance between the upper boundaries of the walls and the calculation area is $H_\mathrm{upper}$, and the distance between the lower boundaries of the walls and the calculation area is $H_\mathrm{lower}$.
The origin of the coordinates is set at the center of the pore, and the calculation area is set as $-d_\mathrm{in}/2 \leq x \leq d_\mathrm{in}/2, -b-H_{\rm lower} \leq y \leq b+H_{\rm upper}$.
For the $i$th calculation area, the Navier-Stokes equation
\begin{align}
  \rho^{(i)} \left[ \frac{\partial \bm{v}^{(i)}}{\partial t} + (\bm{v}^{(i)} \cdot \nabla) \bm{v}^{(i)} \right] = - \nabla p^{(i)} + \mu \nabla^2 \bm{v}^{(i)} + \rho^{(i)} \bm{g},
  \label{eq:Navier_Stokes}
\end{align}
and the incompressible condition
\begin{align}
  \nabla \cdot \bm{v}^{(i)} = 0,
  \label{eq:continuity}
\end{align}
are adopted, where $\rho^{(i)}=\rho^{(i)}(x,y,t)$ is the density, $\bm{v}^{(i)}=\left(v_x^{(i)}(x,y,t),v_y^{(i)}(x,y,t)\right)$ is the velocity, $p^{(i)}=p^{(i)}(x,y,t)$ is the pressure, $\mu$ is the viscosity, and $\bm{g}=(0,-g)$ is the acceleration of gravity.
The normalized concentration $c^{(i)}=c^{(i)}(x,y,t)$ is calculated using the advection-diffusion equation
\begin{align}
  \frac{\partial c^{(i)}}{\partial t} + \nabla \cdot (c^{(i)} \bm{v}^{(i)}) &= D \nabla^2 c^{(i)},
  \label{eq:advection_diffusion}
\end{align}
where $D$ is the diffusion coefficient.
We define the density $\rho^{(i)}$, which depends on the concentration $c^{(i)}$, as
\begin{align}
  \rho^{(i)} = \rho_\mathrm{low} + c^{(i)}(\rho_\mathrm{high} - \rho_\mathrm{low}),
\end{align}
where $\rho_\mathrm{high}$ and $\rho_\mathrm{low}$ are the densities of the higher and lower density fluids, respectively.
The water levels in the $i$th inner container $y_\mathrm{in}^{(i)}(t)$ and the common outer container $y_\mathrm{out}(t)$ are associated with the pressures $p_\mathrm{upper}^{(i)}(t)$ and $p_\mathrm{lower}(t)$ at the upper and lower boundaries of the $i$th calculation area, respectively, as
\begin{subequations}
  \begin{align}
    p_\mathrm{upper}^{(i)} (t) &= \rho_\mathrm{high} g \left(y_\mathrm{in}^{(i)} (t) - b - H_{\mathrm{upper}} \right),
    \label{eq:pressure_couple_i}
    \\
    p_\mathrm{lower} (t) &= \rho_\mathrm{low} g \left(y_\mathrm{out} (t) + b + H_{\mathrm{lower}} \right),
    \label{eq:pressure_couple_o}
  \end{align}
\end{subequations}
where $p_\mathrm{lower}(t)$ is common for all oscillators.
The fluid densities in the inner and outer containers should change with oscillations, but the change is so small that they are assumed to be constants $\rho_\mathrm{high}$ and $\rho_\mathrm{low}$ in Eqs.~(\ref{eq:pressure_couple_i}) and (\ref{eq:pressure_couple_o}), respectively.
The changes in the water levels are obtained from the amount of fluid passing through the pore per unit time $Q^{(i)}(t)$ as
\begin{subequations}
  \begin{align}
    \frac{dy_\mathrm{in}^{(i)}}{dt} &= \frac{Q^{(i)}}{d_\mathrm{in}},
    \label{eq:waterlevel_couple_i}
    \\
    \frac{dy_\mathrm{out}}{dt} &= - \frac{\sum_i Q^{(i)}}{d_\mathrm{out}},
    \label{eq:waterlevel_couple_o}
    \\
    Q^{(i)}(t) &= \int_{-a}^{a} v_y^{(i)} (x, b, t) dx.
    \label{eq:flow}
  \end{align}
\end{subequations}

The boundary conditions are set as follows.
At the boundaries on the walls, the velocity follows the nonslip boundary condition as $\bm{v}^{(i)}=\bm{0}$, and the concentration follows the Neumann boundary condition as $\nabla_\perp c^{(i)} = 0$, where $\nabla_\perp$ denotes the spatial derivative in the direction perpendicular to the boundary.
At the upper and lower boundaries inside the fluid, $\bm{v}^{(i)}$ and $c^{(i)}$ follow the Neumann boundary conditions as $\nabla_\perp \bm{v}^{(i)}=\bm{0}$ and $\nabla_\perp c^{(i)} = 0$, respectively, and the pressure follows Eqs.~(\ref{eq:pressure_couple_i}) and (\ref{eq:pressure_couple_o}).

In the initial state, the higher and lower density fluids are stationary ($\bm{v}^{(i)}=\bm{0}$) and are not mixed with each other ($c^{(i)}=0$ at $y<b$ and $c^{(i)}=1$ at $y \geq b$).
The initial water level in the outer container $y_\mathrm{out}(0)=y_{\mathrm{out},0}$ is fixed in all simulations.
For the single oscillator ($n=1$), the initial water level in the inner container is set from the balance with the hydrostatic pressure by the outer container fluid as $y_\mathrm{in}^{(1)}(0)=b+\left(\rho_\mathrm{low}/\rho_\mathrm{high}\right)\left(y_{\mathrm{out},0}-b\right)$.
For the coupled oscillators ($n\geq2$), the initial water levels in the inner containers $y_\mathrm{in}^{(i)}(0)$ are varied as control parameters.

To numerically solve the Navier-Stokes equation with the incompressible condition in Eqs.~(\ref{eq:Navier_Stokes}) and (\ref{eq:continuity}), the marker-and-cell method was adopted \cite{Mac1,Mac2}, where the Poisson equation obtained from the divergence of Eq.~(\ref{eq:Navier_Stokes}) was calculated for the pressure.
We ignored the gradient of the density, $\nabla \rho^{(i)}$, in the Poisson equation since it is sufficiently small.
The explicit method was adopted for the advection-diffusion equation in Eq.~(\ref{eq:advection_diffusion}).
Each calculation area was divided into $200 \times 240$ meshes, and the spatial mesh size was set to $dx=dy=0.005$.
The time was evolved with the time step $dt=0.0002$.
The parameters were set as follows: $a=0.03$, $b=0.05$, $d_\mathrm{in}=1$, $d_\mathrm{out}=6$, $g=10$, $\mu=1/300$, $D=0.0001$, $\rho_\mathrm{high}=1.2$, $\rho_\mathrm{low}=1$, $H_\mathrm{upper}=H_\mathrm{lower}=0.55$, and $y_{\mathrm{out},0}=10.05$.

\section{simulation results}

We performed the hydrodynamic simulation for the coupled density oscillators with identical intrinsic frequencies and coupling strengths between oscillators for $n=2,3,4$.

\begin{figure*} 
  \begin{center}
    \includegraphics{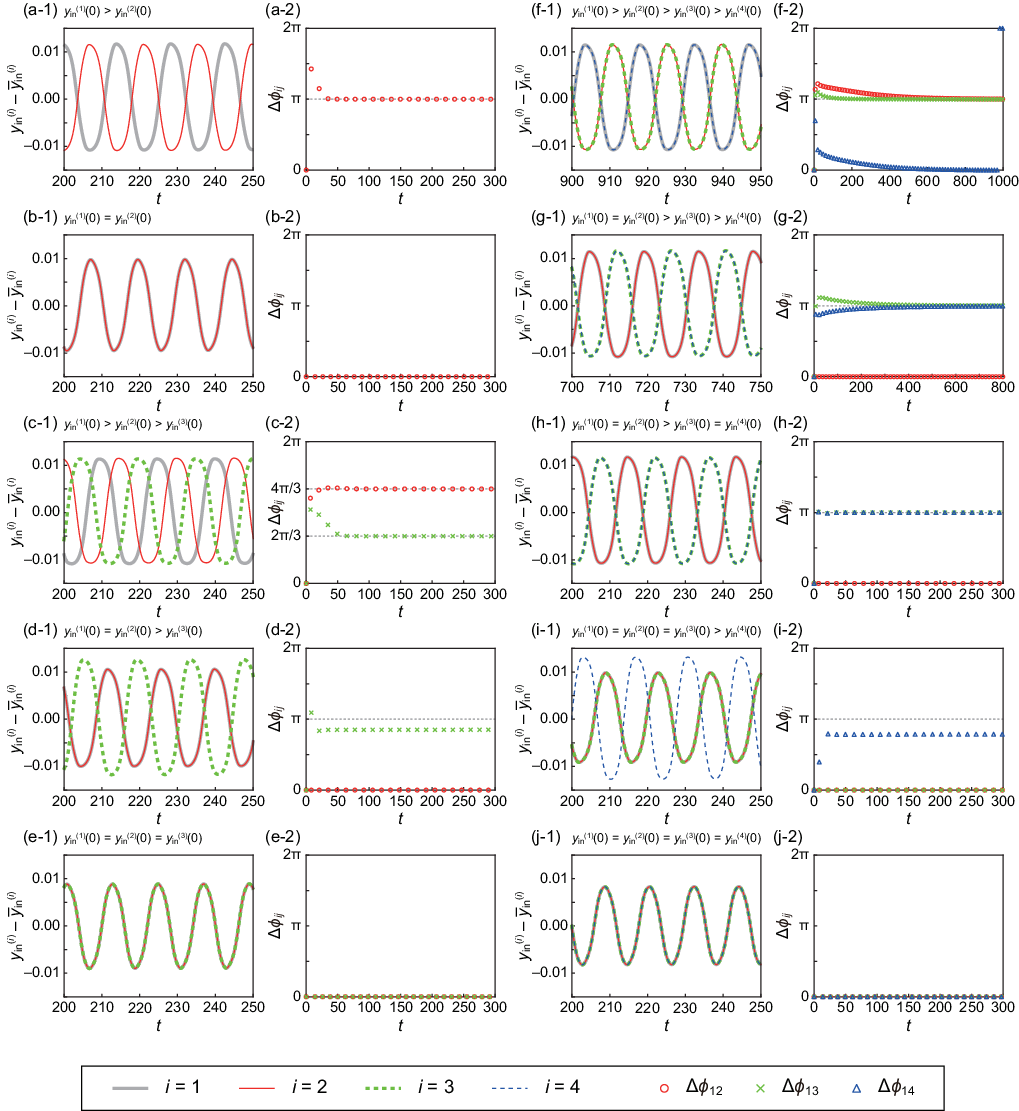}
  \end{center}
  \caption{Synchronization in coupled density oscillators. [(a),(b)] $n=2$, [(c)--(e)] $n=3$, and [(f)--(j)] $n=4$. [(a-1)--(j-1)] Time series of $y_\mathrm{in}^{(i)}-\bar{y}_\mathrm{in}^{(i)}$, where $\bar{y}_\mathrm{in}^{(i)}$ is the time average of $y_\mathrm{in}^{(i)}$. The grey bold solid, red narrow solid, green bold dashed, and blue narrow dashed lines represent $i=1,2,3,$ and $4$, respectively. The symmetry of the initial water level $y_\mathrm{in}^{(i)}(0)$ is shown above each panel. [(a-2)--(j-2)] Corresponding phase differences $\Delta \phi_{ij}=\phi_j-\phi_i$. The red circles, green crosses, and blue triangles represent $\Delta \phi_{12}$, $\Delta \phi_{13}$, and $\Delta \phi_{14}$, respectively.}
  \label{fig:simulation}
\end{figure*}

In the simulation for the two coupled oscillators ($n=2$), an upward flow and a downward flow alternately occurred in each oscillator, and several periods later, the oscillations of flows in both oscillators synchronized; when an upward flow started in one oscillator, a downward flow started in the other oscillator.
The time series of $y_\mathrm{in}^{(1)}$ and $y_\mathrm{in}^{(2)}$ are shown in Fig.~\ref{fig:simulation}(a-1).
To describe the synchronization quantitatively, we calculated the phase difference between two oscillators.
Here, a phase $\phi$ ($0 \leq \phi < 2\pi$) is defined in proportion to time; $\phi=0$ corresponds to the time when an upward flow starts and $\phi=2\pi$ corresponds to when the next upward flow starts [see Fig.~\ref{fig:response}(a)].
The phase difference between two oscillators is represented as $\Delta \phi_{ij}=\phi_j-\phi_i$, where $\phi_i$ is the phase of the $i$th oscillator.
The phase difference $\Delta \phi_{12}$ converged to the constant value $\Delta \phi_{12}=\pi$, i.e., the antiphase mode as shown in Fig.~\ref{fig:simulation}(a-2).

We also realized the in-phase mode ($\Delta \phi_{12} =0$), which has not been observed in experiments, by setting the identical initial conditions for two oscillators.
Due to the symmetric procedure in the numerical calculation, the initial phase difference $\Delta \phi_{12} =0$ was kept, even if it was unstable.
The symmetry of the initial conditions was controlled by the water levels: the antiphase mode appeared from $y_\mathrm{in}^{(1)}(0) \neq y_\mathrm{in}^{(2)}(0)$, and the in-phase mode appeared from $y_\mathrm{in}^{(1)}(0)=y_\mathrm{in}^{(2)}(0)$.

Three coupled oscillators ($n=3$) converged to the three-phase mode with the equivalent phase differences $2\pi/3$ as shown in Fig.~\ref{fig:simulation}(c).
By controlling the symmetry of the initial conditions for a part or all of the oscillators, we realized the partial-in-phase mode from $y_\mathrm{in}^{(1)}(0) = y_\mathrm{in}^{(2)}(0) \neq y_\mathrm{in}^{(3)}(0)$ and the all-in-phase mode from $y_\mathrm{in}^{(1)}(0) = y_\mathrm{in}^{(2)}(0) = y_\mathrm{in}^{(3)}(0)$ as shown in Figs.~\ref{fig:simulation}(d) and \ref{fig:simulation}(e), respectively.

Four coupled oscillators ($n=4$) converged to the 2-2 partial-in-phase mode, where two pairs of in-phase oscillators synchronized with the phase difference $\pi$ as shown in Fig.~\ref{fig:simulation}(f).
It took a longer time to converge to the constant phase differences compared with the antiphase mode ($n=2$) and the three-phase mode ($n=3$).
By controlling the symmetry of the initial conditions for a part or all of the oscillators, we realized the 2-2 partial-in-phase mode [Figs.~\ref{fig:simulation}(g) and \ref{fig:simulation}(h)], the 3-1 partial-in-phase mode [Fig.~\ref{fig:simulation}(i)], and the all-in-phase mode [Fig.~\ref{fig:simulation}(j)].
Here, the 3-1 partial-in-phase mode consists of a set of three in-phase oscillators and an oscillator with a different phase from the other three.

\section{analysis of phase dynamics}

\begin{figure}[tb]
  \begin{center}
    \includegraphics{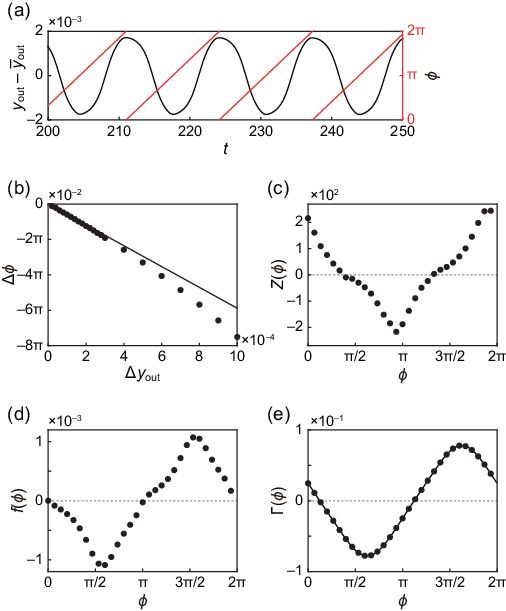}
  \end{center}
  \caption{Phase description of a density oscillator. (a) Definition of the phase $\phi$. The black line shows the time series of $y_\mathrm{out}-\bar{y}_\mathrm{out}$ scaled on the left axis, where $\bar{y}_\mathrm{out}$ is the time average of $y_\mathrm{out}$. The red line shows the corresponding phase $\phi$ scaled on the right axis. (b) Dependence of the phase shift $\Delta \phi$ on the perturbation amplitude $\Delta y_\mathrm{out}$. The perturbation was introduced at $\phi=\pi$ in a cycle, where the phase shift was sufficiently large. The slope of the line was determined by the fitting from five points for $\Delta y_\mathrm{out}\leq0.0001$. (c) Phase sensitivity function $Z (\phi)$ obtained from the perturbation with $\Delta y_\mathrm{out}=0.0001$. (d) Periodic perturbation $f(\phi)$ to the water level in the outer container. (e) Phase coupling function $\Gamma (\phi)$. The points represent the discrete data calculated from (c) and (d). The line shows the fitting curve by the Fourier series up to sixth order.}
  \label{fig:response}
\end{figure}

We analyzed the stabilities of the synchronization modes from the viewpoint of phase dynamics.
We first investigated the single density oscillator subjected to a perturbation to measure the phase response for the limit-cycle oscillation ($n=1$).
Figure~\ref{fig:response}(a) shows the time series of the outer water level $y_\mathrm{out}$ without perturbation and the corresponding phase $\phi$.
The perturbation of the outer water level $\Delta y_\mathrm{out}$ is introduced at $t=t_0$, and the time evolution of $y_\mathrm{out}$ with the perturbation is given instead of Eq.~(\ref{eq:waterlevel_couple_o}) as
\begin{align}
  \frac{dy_\mathrm{out}}{dt} &= - \frac{Q^{(1)}}{d_\mathrm{out}} + \Delta y_\mathrm{out} \delta (t-t_0).
  \label{eq:waterlevel_couple_p}
\end{align}
The phase shift $\Delta \phi$ is determined by subtracting the phase $\phi_0$ in the unperturbed system from the phase $\phi^\prime$ in the perturbed system after sufficiently long time as
\begin{align}
  \Delta \phi = \phi^\prime - \phi_0.
  \label{eq:phase_shift}
  \end{align}
In the simulation, the perturbation was introduced 15 periods after the initial state, and the phase shift was measured 15 periods after the perturbation was introduced to ensure sufficient relaxation.
The phase response curve $\Delta \phi = \Delta \phi (\phi,\Delta y_\mathrm{out})$ represents the phase shift as a function of the phase $\phi$ at which the perturbation is introduced and the perturbation amplitude $\Delta y_\mathrm{out}$.
For a sufficiently small perturbation, the phase response curve is expected to be linear to the perturbation amplitude $\Delta y_\mathrm{out}$ as
\begin{align}
  \Delta \phi (\phi,\Delta y_\mathrm{out}) = \Delta y_\mathrm{out} Z(\phi),
  \label{eq:linear}
  \end{align}
where $Z(\phi)$ is the phase sensitivity function.
Figure~\ref{fig:response}(b) shows the phase shift $\Delta \phi$ plotted against the perturbation amplitude $\Delta y_\mathrm{out}$, and the linear region where Eq.~(\ref{eq:linear}) holds was confirmed below $\Delta y_\mathrm{out} \sim 0.0002$.
Then, we measured the phase sensitivity function $Z(\phi)$ with the perturbation amplitude $\Delta y_\mathrm{out} = 0.0001$ as shown in Fig.~\ref{fig:response}(c).
The phase sensitivity function $Z(\phi)$ is positive, and the phase is preceded at $0 \lesssim \phi \lesssim \pi/3,4\pi/3 \lesssim \phi \lesssim 2\pi$.
In contrast, $Z(\phi)$ is negative, and the phase is delayed at $\pi/3 \lesssim \phi \lesssim 4\pi/3$.
These behaviors could be interpreted as follows.
Due to an increase in the outer water level by the perturbation, the upward or downward flow ran out earlier for $Z(\phi)>0$, whereas it kept longer for $Z(\phi)<0$, compared to those without the perturbation.
The change in the sign of $Z(\phi)$ might be associated with the acceleration of the outer water level $d^2y_\mathrm{out}/dt^2$ ($=\omega df(\phi)/d\phi$) [see Fig.~\ref{fig:response}(d)].
An increase in the outer water level by the perturbation can precede the phase if the upward flow is getting strong or the downward flow is getting weak.
In contrast, it can delay the phase if the upward flow is getting weak or the downward flow is getting strong.

\begin{figure}[tb]
  \begin{center}
    \includegraphics{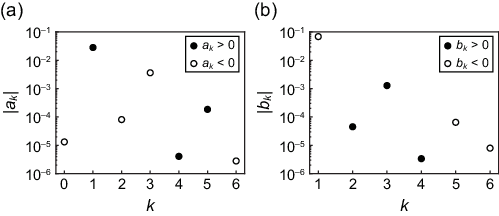}
  \end{center}
  \caption{Fourier components of $\Gamma(\phi)$ shown in Fig.~\ref{fig:response}(e). (a) Fourier cosine coefficient $a_k$. (b) Fourier sine coefficient $b_k$. The signs of $a_k$ and $b_k$ are represented by solid circles (positive) and open circles (negative).}
  \label{fig:components}
\end{figure}

The phase dynamics of $n$ weakly coupled identical oscillators can be analyzed from the phase response of a single oscillator under the small perturbation based on the phase reduction theory \cite{Kuramoto}.
The dynamics of the $i$th oscillator under the small perturbation from other oscillators is described as
\begin{align}
  \frac{d\phi_i}{dt} = \omega + Z(\phi_i) \sum_{j \ne i}f(\phi_j),
  \label{eq:phase}
  \end{align}
where $\omega$ is the intrinsic frequency of the limit-cycle oscillation, and $f(\phi_i)=-Q^{(1)}(\phi_i)/d_\mathrm{out}$ is the periodic perturbation by the $i$th oscillator to $y_\mathrm{out}$ as shown in Fig.~\ref{fig:response}(d).
The periodic time series of $Q^{(1)}$ was obtained from the time series 15 periods after the initial state for $n=1$.
Equation~(\ref{eq:phase}) is approximated by taking the time average of the second term on the right-hand side over a period as
\begin{align}
  \frac{d\phi_i}{dt} &= \omega + \sum_{j \ne i}\Gamma(\Delta\phi_{ij}),
  \label{eq:average}
  \end{align}
where $\Gamma(\phi)$ is the phase coupling function:
\begin{align}
  \Gamma(\phi) = \frac{1}{2\pi} \int_{0}^{2\pi}Z\left(\theta\right)f\left(\phi+\theta\right)d\theta.
  \label{eq:gamma}
  \end{align}
Figure~\ref{fig:response}(e) shows the phase coupling function $\Gamma(\phi)$ calculated from $Z(\phi)$ and $f(\phi)$ in Fig.~\ref{fig:response}(c) and \ref{fig:response}(d), respectively.
$\Gamma(\phi)$ looks like a sinusoidal function although $Z(\phi)$ and $f(\phi)$ seem to include higher harmonics.
This reason is discussed in detail in Appendix A.
Here, we consider the Fourier series
\begin{align}
  \Gamma(\phi)=\frac{a_0}{2}+\sum_{k=1}^{\infty}\left(a_k \cos(k\phi)+b_k \sin(k\phi)\right),
  \label{eq:fourier}
\end{align}
where $a_k$ and $b_k$ are the Fourier cosine and sine coefficients of $\Gamma(\phi)$, respectively.
The coefficients $a_k$ and $b_k$ for $k\leq6$ calculated from the discrete points in Fig.~\ref{fig:response}(e) are shown in Fig.~\ref{fig:components}.
For the following calculation including $\Gamma(\phi)$, we use the Fourier series expanded up to sixth order.
The phase coupling function represented by the Fourier series is also shown with a line in Fig.~\ref{fig:response}(e), which well fits the discrete points.

\begin{figure}[tb]
  \begin{center}
    \includegraphics{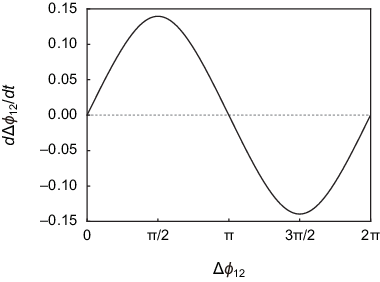}
  \end{center}
  \caption{Phase portrait in the ($\Delta\phi_{12}, d\Delta\phi_{12}/dt$) plane for two coupled oscillators. It indicates that the antiphase mode for $\Delta\phi_{12}=\pi$ is stable, whereas the in-phase mode for $\Delta\phi_{12}=0$ is unstable.}
  \label{fig:2_coupled}
\end{figure}

The dynamics of the phase difference $\Delta\phi_{12}$ in two coupled oscillators is derived as
\begin{align}
  \frac{d\Delta\phi_{12}}{dt} = \Gamma(-\Delta\phi_{12}) - \Gamma(\Delta\phi_{12}).
  \label{eq:phaseeq_2}
\end{align}
The time derivative of $\Delta\phi_{12}$ is expressed as a function of $\Delta\phi_{12}$, and the phase portrait is obtained as shown in Fig.~\ref{fig:2_coupled}.
The fixed points, where $d\Delta\phi_{12}/dt=0$ holds, are $\Delta\phi_{12}=0$ and $\pi$.
The derivative $d\Delta\phi_{12}/dt$ is positive at $0<\Delta\phi_{12}<\pi$ and negative at $\pi<\Delta\phi_{12}<2\pi$.
Therefore, the antiphase mode corresponding to $\Delta\phi_{12}=\pi$ is stable, whereas the in-phase mode corresponding to $\Delta\phi_{12}=0$ is unstable.

\begin{figure}[tb]
  \begin{center}
    \includegraphics{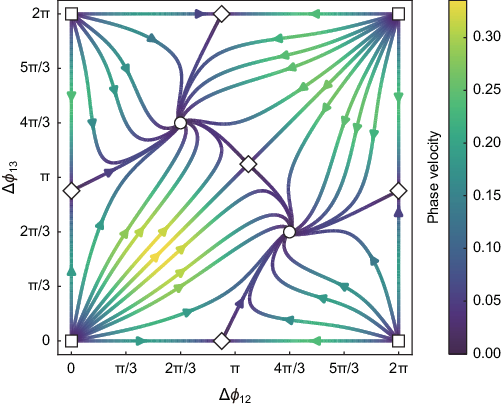}
  \end{center}
  \caption{Vector field of ($d\Delta \phi_\mathrm{12}/dt, d\Delta \phi_\mathrm{13}/dt$) in the ($\Delta \phi_\mathrm{12}, \Delta \phi_\mathrm{13}$) plane for three coupled oscillators. The unstable star nodes (square), the unstable saddle points (diamond), and the stable spirals (circle) represent the all-in-phase, partial-in-phase, and three-phase modes, respectively. The phase velocity $\sqrt{\left( d\Delta\phi_{12}/dt \right)^2 + \left( d\Delta\phi_{13}/dt \right)^2}$ in the ($\Delta \phi_\mathrm{12}, \Delta \phi_\mathrm{13}$) plane is shown as a color gradient.}
  \label{fig:3_coupled}
\end{figure}

The dynamics of the phase differences $\Delta\phi_{12}$ and $\Delta\phi_{13}$ in three coupled oscillators are derived as
\begin{subequations}
  \begin{align}
    \frac{d\Delta\phi_{12}}{dt} &= \Gamma(-\Delta\phi_{12}) +\Gamma(\Delta\phi_{13}-\Delta\phi_{12}) \nonumber\\ &\quad - \Gamma(\Delta\phi_{12}) - \Gamma(\Delta\phi_{13}),
    \label{eq:phaseeq_3a}
    \\
    \frac{d\Delta\phi_{13}}{dt} &= \Gamma(-\Delta\phi_{13}) +\Gamma(\Delta\phi_{12}-\Delta\phi_{13}) \nonumber\\ &\quad - \Gamma(\Delta\phi_{12}) - \Gamma(\Delta\phi_{13}).
    \label{eq:phaseeq_3b}
    \end{align}
    \label{eq:phaseeq_3ab}
  \end{subequations}
The stabilities of the phase differences are visualized on the vector field in the ($\Delta\phi_{12}, \Delta\phi_{13}$) plane as shown in Fig.~\ref {fig:3_coupled}.
There are three synchronization modes satisfying $d\Delta\phi_{12}/dt=d\Delta\phi_{13}/dt=0$: (i) the all-in-phase mode for $(\Delta\phi_{12},\Delta\phi_{13})=(0,0)$, which is an unstable star node; (ii) the partial-in-phase mode for $(\Delta\phi_{12},\Delta\phi_{13})=(\alpha,0),(0,\alpha),(2\pi-\alpha,2\pi-\alpha)$, which are unstable saddle points; and (iii) the three-phase mode for $(\Delta\phi_{12},\Delta\phi_{13})=(2\pi/3,4\pi/3),(4\pi/3,2\pi/3)$, which are stable spirals.
Here, $\alpha$ satisfies $\Gamma(\alpha)=2\Gamma(-\alpha)-\Gamma(0)$ and $0<\alpha<2\pi$.

\begin{figure}[tb]
  \begin{center}
    \includegraphics{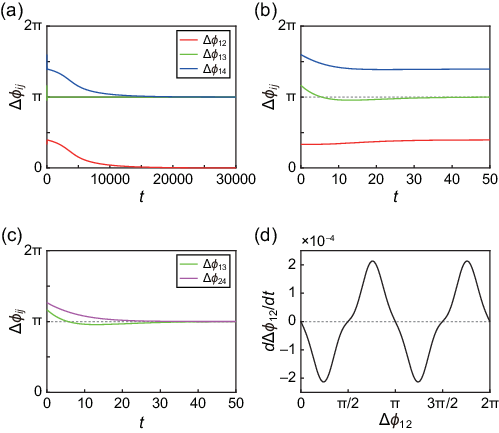}
  \end{center}
  \caption{Behavior of the phase differences $\Delta \phi_{12}$, $\Delta \phi_{13}$, and $\Delta \phi_{14}$ in four coupled oscillators. (a) Time series of the phase differences from the initial state $(\Delta\phi_{12},\Delta\phi_{13},\Delta\phi_{14})=(\pi/3,7\pi/6,8\pi/5)$. (b) Magnified image for $0\leq t \leq 50$ in (a). (c) $\Delta \phi_{13}$ and $\Delta \phi_{24}$ during the same time range as in (b). The red, green, blue, and magenta lines represent $\Delta \phi_{12}$, $\Delta \phi_{13}$, $\Delta \phi_{14}$, and $\Delta \phi_{24}$, respectively. (d) Phase portrait in the ($\Delta\phi_{12}, d\Delta\phi_{12}/dt$) plane for the four coupled oscillators with two antiphase pairs $(\Delta\phi_{12},\Delta\phi_{13},\Delta\phi_{14})=(\Delta\phi_{12},\pi,\Delta\phi_{12}+\pi)$. It indicates that the 2-2 partial-in-phase mode for $\Delta\phi_{12}=0$ and $\pi$ is stable, whereas the four-phase mode for $\Delta\phi_{12}=\pi/2$ and $3\pi/2$ is unstable.}
  \label{fig:4_coupled}
\end{figure}

The dynamics of the phase differences $\Delta\phi_{12}$, $\Delta\phi_{13}$, and $\Delta\phi_{14}$ in four coupled oscillators are derived as
\begin{subequations}
  \begin{align}
    \frac{d\Delta\phi_{12}}{dt} 
    &= \Gamma(-\Delta\phi_{12}) +\Gamma(\Delta\phi_{13}-\Delta\phi_{12}) \nonumber\\
    &\quad +\Gamma(\Delta\phi_{14}-\Delta\phi_{12}) \nonumber\\ 
    &\quad -\Gamma(\Delta\phi_{12}) -\Gamma(\Delta\phi_{13}) -\Gamma(\Delta\phi_{14}),
    \label{eq:phaseeq_4a}
    \\
    \frac{d\Delta\phi_{13}}{dt} 
    &= \Gamma(-\Delta\phi_{13}) +\Gamma(\Delta\phi_{12}-\Delta\phi_{13}) \nonumber\\
    &\quad +\Gamma(\Delta\phi_{14}-\Delta\phi_{13}) \nonumber\\ 
    &\quad -\Gamma(\Delta\phi_{12}) -\Gamma(\Delta\phi_{13}) -\Gamma(\Delta\phi_{14}),
    \label{eq:phaseeq_4b}
    \\
    \frac{d\Delta\phi_{14}}{dt} 
    &= \Gamma(-\Delta\phi_{14}) +\Gamma(\Delta\phi_{12}-\Delta\phi_{14}) \nonumber\\
    &\quad +\Gamma(\Delta\phi_{13}-\Delta\phi_{14}) \nonumber\\ 
    &\quad -\Gamma(\Delta\phi_{12}) -\Gamma(\Delta\phi_{13}) -\Gamma(\Delta\phi_{14}).
    \label{eq:phaseeq_4c}
  \end{align}
  \label{eq:phaseeq_4abc}
\end{subequations}
There are four synchronization modes satisfying $d\Delta\phi_{12}/dt=d\Delta\phi_{13}/dt=d\Delta\phi_{14}/dt=0$: (i) the all-in-phase mode for $(\Delta\phi_{12},\Delta\phi_{13},\Delta\phi_{14})=(0,0,0)$, (ii) the 3-1 partial-in-phase mode for $(\Delta\phi_{12},\Delta\phi_{13},\Delta\phi_{14})=(\beta,0,0),(0,\beta,0),(0,0,\beta),(2\pi-\beta,2\pi-\beta,2\pi-\beta)$, (iii) the 2-2 partial-in-phase mode for $(\Delta\phi_{12},\Delta\phi_{13},\Delta\phi_{14})=(0,\pi,\pi),(\pi,0,\pi),(\pi,\pi,0)$, and (iv) the four-phase mode for $(\Delta\phi_{12},\Delta\phi_{13},\Delta\phi_{14})=(\pi/2,\pi,3\pi/2),(\pi/2,3\pi/2,\pi),(\pi,\pi/2,3\pi/2),(\pi,3\pi/2,\allowbreak\pi/2),(3\pi/2,\pi/2,\pi),(3\pi/2,\pi,\pi/2)$.
Here, $\beta$ satisfies $\Gamma(\beta)=3\Gamma(-\beta)-2\Gamma(0)$ and $0<\beta<2\pi$.
Instead of visualizing the three-dimensional dynamics of the phase differences, we investigate the time variation of the phase differences from arbitrary initial states.
Equations~(\ref{eq:phaseeq_4a})--(\ref{eq:phaseeq_4c}) were discretized with the Euler method, and the time evolutions of $\Delta\phi_{12},\Delta\phi_{13}$, and $\Delta\phi_{14}$ were calculated with the time step $dt=0.01$.
The numerical calculation indicated that the system converged to the 2-2 partial-in-phase mode as shown in Fig.~\ref{fig:4_coupled}(a), where the magnified image around the initial state is shown in Fig.~\ref{fig:4_coupled}(b).
Figure~\ref{fig:4_coupled}(c) shows the phase differences $\Delta\phi_{13}$ and $\Delta\phi_{24}$ during the same time range as in (b).
Two antiphase pairs were rapidly formed in the time scale of $t\sim 10^1$, whereas the system converged to the 2-2 partial-in-phase mode in the time scale of $t\sim 10^4$.
Thus we consider the phase dynamics of the two antiphase pairs $(\Delta\phi_{12},\Delta\phi_{13},\Delta\phi_{14})=(\Delta\phi_{12},\pi,\Delta\phi_{12}+\pi)$.
Equations~(\ref{eq:phaseeq_4a})--(\ref{eq:phaseeq_4c}) are reduced to
\begin{subequations}
  \begin{align}
    \frac{d\Delta\phi_{12}}{dt} &= \Gamma(-\Delta\phi_{12})-\Gamma(\Delta\phi_{12}) \nonumber\\ &\quad +\Gamma(\pi-\Delta\phi_{12})-\Gamma(\Delta\phi_{12}+\pi),
    \label{eq:phaseeq_4d}
    \\
    \frac{d\Delta\phi_{13}}{dt} &= \frac{d\Delta\phi_{24}}{dt}=0.
    \label{eq:phaseeq_4e}
  \end{align}
  \label{eq:phaseeq_4de}
\end{subequations}
The phase portrait in the ($\Delta\phi_{12}, d\Delta\phi_{12}/dt$) plane for the four coupled oscillators with two antiphase pairs is shown in Fig.~\ref{fig:4_coupled}(d).
The fixed points, where $d\Delta\phi_{12}/dt=0$ holds, are $\Delta\phi_{12}=0,\pi/2,\pi$, and $3\pi/2$.
The derivative $d\Delta\phi_{12}/dt$ is positive at $\pi/2<\Delta\phi_{12}<\pi,3\pi/2<\Delta\phi_{12}<2\pi$ and negative at $0<\Delta\phi_{12}<\pi/2,\pi<\Delta\phi_{12}<3\pi/2$.
Therefore, the 2-2 partial-in-phase mode corresponding to $\Delta\phi_{12}=0$ and $\pi$ is stable, whereas the four-phase mode corresponding to $\Delta\phi_{12}=\pi/2$ and $3\pi/2$ is unstable.

We compare the analysis of the phase dynamics with the simulation results in Sec. III.
The antiphase ($n=2$), three-phase ($n=3$), and 2-2 partial-in-phase ($n=4$) modes obtained from the simulations with asymmetric initial conditions were found to be stable, whereas the other modes obtained from the simulations with symmetric initial conditions for a part or all of the oscillators were found to be unstable.

The difference in the period between different synchronization modes in the simulation results can be explained based on the phase reduction theory.
Assuming the synchronization with the constant period $T$ and integrating the phase equation in Eq.~(\ref{eq:average}) over $T$, we obtain
\begin{align}
  T=\frac{1}{1+\frac{\sum_{j \ne i}\Gamma(\Delta\phi_{ij})}{2\pi}T_0} T_0,
  \label{eq:period}
\end{align}
where $T_0=2\pi/\omega$ is an intrinsic period.
It is found that the synchronization period $T$ is longer or shorter than the intrinsic period $T_0$ for negative or positive $\sum_{j \ne i}\Gamma(\Delta\phi_{ij})$, respectively.
The periods obtained from the simulation and those estimated by Eq.~(\ref{eq:period}) are shown in Table~\ref{tab:period}.
Simulation results indicate that synchronization periods are longer or shorter than the intrinsic period for $\sum_{j \ne i}\Gamma(\Delta\phi_{ij})<0$ or $\sum_{j \ne i}\Gamma(\Delta\phi_{ij})>0$, respectively, which qualitatively agrees with the evaluation by Eq.~(\ref{eq:period}).
The quantitative difference in periods between the simulation and the estimation could be attributed to the following two factors.
The first is a slight difference in amplitudes of coupled oscillators from the single oscillator in the simulation as shown in Table~\ref{tab:period}.
They are assumed to be equal in the phase reduction theory.
The second is the time averaging from Eq.~(\ref{eq:phase}) to Eq.~(\ref{eq:average}) in the phase reduction.
\begin{table*}
  \caption{\label{tab:period}Period and amplitude of the oscillation in synchronization. The first, second, third, fourth, fifth, and sixth columns show the number of coupled oscillators, the synchronization mode, the period obtained from the simulation, the period estimated by Eq.~(\ref{eq:period}) with $T_0$ and $\Gamma(\phi)$ obtained from the simulation, the sign of $\sum_{j \neq i}\Gamma(\Delta\phi_{ij})$, and the amplitude obtained from the simulation, respectively.}
  \begin{ruledtabular}
  \begin{tabular}{cccccc}
    Number of & Synchronization & Period by & Estimated & Sign of & Amplitude\\
    oscillators& mode & simulation & period & $\sum_{j \neq i}\Gamma(\Delta\phi_{ij})$ & by simulation\\ \hline
    1  &  & 13.2 &  &  & 0.0104\\
    \addlinespace[4mm]
    2  & antiphase & 14.4 & 13.9 & negative & 0.0112 \\
    2  & in-phase & 12.5 & 12.5 & positive & 0.0096 \\
    \addlinespace[4mm]
    3  & three-phase & 15.3 & 14.3 & negative & 0.0111\\
    3  & partial-in-phase & 14.1 & 13.7 & negative & 0.0102\,($i=1,2$),\: 0.0121\,($i=3$)\\
    3  & all-in-phase & 12.1 & 12.0 & positive & 0.0089\\
    \addlinespace[4mm]
    4  & 2-2 partial-in-phase\footnotemark[1] & 14.4 & 13.9 & negative & 0.0112\\
    4  & 3-1 partial-in-phase & 13.8 & 13.2 & negative & 0.0095\,($i=1,2,3$),\: 0.0129\,($i=4$)\\
    4  & all-in-phase & 11.8 & 11.4 & positive & 0.0082\\
  \end{tabular}
  \end{ruledtabular}
  \footnotetext[1]{Among the three simulation results in Figs.~\ref{fig:simulation}(f)--\ref{fig:simulation}(h), the simulation result of Fig.~\ref{fig:simulation}(h) is adopted.}
\end{table*}

\section{linear stability analysis}

A linear stability analysis is applied to the fixed points of the phase differences between coupled oscillators in Eqs.~(\ref{eq:phaseeq_2}), (\ref{eq:phaseeq_3ab}), and (\ref{eq:phaseeq_4abc}) for $n=2,3,$ and $4$, respectively.
We describe the stabilities of the fixed points generically with the Fourier components $a_k$ and $b_k$ of the phase coupling function $\Gamma(\phi)$ in Eq.~(\ref{eq:fourier}).

For the two coupled oscillators, we consider a fixed point $\Delta \phi_{12}^{*}$ and the perturbation $\eta_{12}(t)$ from a fixed point, which is defined as $\eta_{12}(t)=\Delta \phi_{12}(t)-\Delta \phi_{12}^{*}$.
The perturbation approximately follows
\begin{align}
  \frac{d\eta_{12}}{dt}= \lambda \eta_{12},
\end{align}
where $\lambda$ is the derivative of the right-hand side in Eq.~(\ref{eq:phaseeq_2}) with respect to $\Delta \phi_{12}$ at the fixed point $\Delta \phi_{12}=\Delta \phi_{12}^{*}$.

For the three coupled oscillators, we consider a fixed point ($\Delta \phi_{12}^{*},\Delta \phi_{13}^{*}$) and the perturbation ($\eta_{12},\eta_{13}$) from a fixed point, which is defined as $(\eta_{12},\eta_{13})=(\Delta \phi_{12},\Delta \phi_{13})-(\Delta \phi_{12}^{*},\Delta \phi_{13}^{*})$.
The perturbation approximately follows
\begin{align}
  \frac{d}{dt}
  \begin{pmatrix}
    \eta_{12} \\
    \eta_{13} 
  \end{pmatrix}
  =A
  \begin{pmatrix}
    \eta_{12} \\
    \eta_{13}
  \end{pmatrix},
\end{align}
where $A$ is the Jacobian matrix of the right-hand sides in Eqs.~(\ref{eq:phaseeq_3a}) and (\ref{eq:phaseeq_3b}) at the fixed point ($\Delta \phi_{12},\Delta \phi_{13}$)=($\Delta \phi_{12}^{*},\Delta \phi_{13}^{*}$).

For the four coupled oscillators, we consider a fixed point ($\Delta \phi_{12}^{*},\Delta \phi_{13}^{*},\Delta \phi_{14}^{*}$) and the perturbation ($\eta_{12},\eta_{13},\eta_{14}$) from a fixed point, which is defined as $(\eta_{12},\eta_{13},\eta_{14})=(\Delta \phi_{12},\Delta \phi_{13},\Delta \phi_{14})-(\Delta \phi_{12}^{*},\Delta \phi_{13}^{*},\Delta \phi_{14}^{*})$.
The perturbation approximately follows
\begin{align}
  \frac{d}{dt}
  \begin{pmatrix}
    \eta_{12} \\
    \eta_{13} \\
    \eta_{14}
  \end{pmatrix}
  =A
  \begin{pmatrix}
    \eta_{12} \\
    \eta_{13} \\
    \eta_{14}
  \end{pmatrix},
  \label{eq:appendix}
\end{align}
where $A$ is the Jacobian matrix of the right-hand sides in Eqs.~(\ref{eq:phaseeq_4a})--(\ref{eq:phaseeq_4c}) at the fixed point ($\Delta \phi_{12},\Delta \phi_{13},\Delta \phi_{14}$)=($\Delta \phi_{12}^{*},\Delta \phi_{13}^{*},\Delta \phi_{14}^{*}$).

The stability of the fixed point can be evaluated from the eigenvalues $\lambda$ of the Jacobian matrix $A$, which indicate the growth rates of the perturbation.
In our simulation, the first Fourier components $a_1$ and $b_1$ are much larger than the higher-order components ($k \geq 2$) [see Fig.~\ref{fig:components}]. Thus $a_1$ and $b_1$ most contribute to the stability if they are included in the real parts of the eigenvalues.
Therefore, we discuss the stability only with $a_1$ and $b_1$ in the eigenvalues.
The results of the linear stability analysis for various synchronization modes are shown in Table~\ref{tab:stability}, where the eigenvalues, the types of the fixed points, and the stabilities evaluated from the numerical result of $a_1$ and $b_1$ are indicated.
We obtained $\alpha = \arctan\left(-6a_1b_1 / (a_1^2-9b_1^2)\right)$ for the partial-in-phase mode ($n=3$) and $\beta = \arctan\left(-4a_1b_1 / (a_1^2-4b_1^2)\right)$ for the 3-1 partial-in-phase mode ($n=4$) only with $a_1$ and $b_1$.
The results of the linear stability analysis suggest that if the antiphase mode is stable in the two coupled oscillators (i.e., $b_1$ is negative), the three-phase mode is also stable in the three coupled oscillators as long as the first Fourier sine coefficient $b_1$ is sufficiently larger than the higher-order ones.

The eigenvalues are zero for the four-phase and 2-2 partial-in-phase modes for $n=4$ due to the neglect of the higher-order components.
In this case, the higher-order components should be considered to evaluate the stability.
With the numerical result of the Fourier components for $k \leq 6$, it was found that the 2-2 partial-in-phase mode is stable, whereas the four-phase mode is unstable.
Here, the eigenvalues including the higher-order components are shown in Appendix B.
The results of the linear stability analysis are consistent with the stabilities obtained in Sec. IV.
\begin{table*}
  \caption{\label{tab:stability}Linear stability analysis for the fixed point of the phase difference in coupled oscillators only with the first Fourier components $a_1$ and $b_1$. The first, second, third, fourth, and fifth columns show the number of coupled oscillators, the synchronization mode, the eigenvalue $\lambda$, the type of the fixed point, and the stability of the fixed point evaluated from the numerical result of $a_1$ and $b_1$, respectively.}
  \begin{ruledtabular}
  \begin{tabular}{ccccc}
    Number of & Synchronization mode & Eigenvalue & Fixed point type & Linear\\
    oscillators & & & & stability\\ \hline
    2  & antiphase & $2b_1$ & node & stable\\
    2  & in-phase & $-2b_1$ & node & unstable\\
    \addlinespace[4mm]
    3  & three-phase & $3\left(b_1\pm ia_1\right)/2$ & spiral & stable\\
    3  & partial-in-phase & $3b_1, \, -9b_1(a_1^2+b_1^2)/(a_1^2+9b_1^2)$ & saddle & unstable\\
    3  & all-in-phase & $-3b_1 \, \text{(duplicate)}$ & node & unstable\\
    \addlinespace[4mm]
    4  & four-phase & $0\footnotemark[1], \,  2\left(b_1\pm ia_1\right)$ & &\\
    4  & 2-2 partial-in-phase & $0\footnotemark[1] \, \text{(duplicate)}, \, 4b_1$ & &\\
    4  & 3-1 partial-in-phase & $4b_1, \, -8b_1(a_1^2+b_1^2)/(a_1^2+4b_1^2) \, \text{(duplicate)}$ & saddle &unstable\\
    4  & all-in-phase & $-4b_1 \, \text{(triplicate)}$ & node & unstable\\
  \end{tabular}
  \end{ruledtabular}
  \footnotetext[1]{$a_1$ or $b_1$ is not included in the eigenvalue.}
\end{table*}

Okuda applied a linear stability analysis to the symmetric cluster states with equivalent phase differences in globally coupled oscillators, where the eigenvalue was described with the Fourier components \cite{Okuda}.
The symmetric cluster states, where each cluster consists of an equal number of oscillators, include the synchronization modes obtained in our study, except for the partial-in-phase ($n=3$) and 3-1 partial-in-phase ($n=4$) modes.
The eigenvalues for the symmetric cluster states obtained in our study correspond to those described by Okuda.

\section{discussion}

We compare our simulation results with experimental results reported in previous studies.
In the experiments of two coupled oscillators, the antiphase mode was observed \cite{Nakata,Yoshikawa,Kano2,Horie}.
In the experiments of three coupled oscillators, the three-phase mode was observed \cite{Yoshikawa,Kano2,Miyakawa2,Horie}.
In our simulation, the antiphase and three-phase modes were obtained for $n=2$ and $n=3$, respectively, which agreed with the experimental observations.
In the experiment of four coupled oscillators, the four-phase mode was observed for the stronger coupling, and the 2-2 partial-in-phase mode was observed for the weaker coupling \cite{Miyakawa2}.
Since the weak coupling was assumed in our study, our simulation for $n=4$, where the 2-2 partial-in-phase mode appeared, agreed with this experimental observation.

In addition to the 2-2 partial-in-phase mode, the 2-1-1 partial-in-phase mode was also observed for the weaker coupling \cite{Horie}.
Here, the 2-1-1 partial-in-phase mode consists of a pair of in-phase oscillators and the other two oscillators with different phases.
This mode did not appear within our simulation and analysis based on the phase reduction theory.
In our study, we consider the interaction only through the common pressure among coupled identical oscillators.
In the experiment, the hydrodynamic interaction could also work between neighboring oscillators depending on the arrangement of the inner containers.
In addition, the oscillators in the experimental system are not exactly identical due to the experimental error or the external noise.
These factors could be related to the emergence of the 2-1-1 partial-in-phase mode only in the experiment.

Regarding the relation between the intrinsic period and the synchronization period in two coupled oscillators, it was reported that the period in the antiphase mode was almost twice the intrinsic period in the experiment \cite{Miyakawa1}.
It was also reported that the period in the in-phase mode was shorter than the intrinsic period and decreased with an increase in the number of coupled oscillators in the numerical calculation using the model with the ordinary differential equations \cite{Kenfack1}.
Our simulation results and the evaluation by Eq.~(\ref{eq:period}) qualitatively agreed with the results in these studies.
To quantitatively compare our simulation result with the experimental one, the dependence of the synchronization period on the parameters such as the intrinsic period or coupling strength needs to be investigated, which is left as future work.

\begin{figure}[tb]
  \begin{center}
    \includegraphics{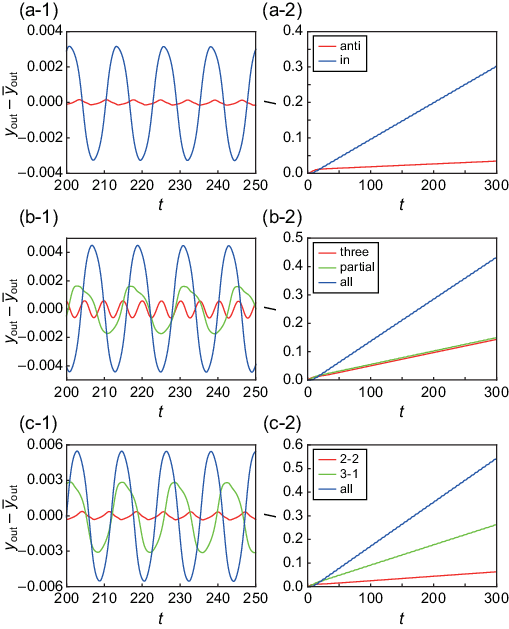}
  \end{center}
  \caption{Comparison of the water level in the outer container $y_\mathrm{out}$ among the synchronization modes for (a) $n=2$, (b) $n=3$, and (c) $n=4$. [(a-1)--(c-1)] Time series of $y_\mathrm{out}-\bar{y}_\mathrm{out}$, where $\bar{y}_\mathrm{out}$ is the time average of $y_\mathrm{out}$. [(a-2)--(c-2)] Time series of the total distance of the change in the water level $l$. Among the three simulation results indicating the 2-2 partial-in-phase mode in Figs.~\ref{fig:simulation}(f)--\ref{fig:simulation}(h), the simulation result of Fig.~\ref{fig:simulation}(h) is adopted.}
  \label{fig:y_out}
\end{figure}

We finally discuss the change in the water level as the criterion for selection of the synchronization mode.
In the antiphase ($n=2$), three-phase ($n=3$), and 2-2 partial-in-phase ($n=4$) modes, which are stable, the changes in $y_\mathrm{out}$ are expected to be canceled out among the oscillators with different phases.
We confirm the correlation between the change in $y_\mathrm{out}$ and the stability of the synchronization modes obtained in the simulation.
Figure~\ref{fig:y_out} shows the time series of $y_\mathrm{out}$ in Figs.~\ref{fig:y_out}(a-1)--\ref{fig:y_out}(c-1) and the total distances of the changes in the water levels $l(t)$ in Figs.~\ref{fig:y_out}(a-2)--\ref{fig:y_out}(c--2) for the $n$ coupled oscillators, where $l(t)$ is defined as
\begin{align}
  l(t)=\int_0^t \frac{\left\lvert\sum_i Q^{(i)}(t^\prime)\right\rvert}{d_\mathrm{out}}dt^\prime.
\end{align} 
The slope of $l$ was evaluated as the rate of the absolute change in $y_\mathrm{out}$.
For the two coupled oscillators ($n=2$), the slope of $l$ in the antiphase mode, which was stable, was much smaller than that in the in-phase mode, which was unstable, as shown in Fig.~\ref{fig:y_out}(a-2).
For the three coupled oscillators ($n=3$), the slope of $l$ in the three-phase mode, which was stable, was also smaller than those in the partial- and all-in-phase modes, which were unstable, as shown in Fig.~\ref{fig:y_out}(b-2).
For the four coupled oscillators ($n=4$), the slope of $l$ in the 2-2 partial-in-phase mode, which was stable, was also smaller than those in the 3-1 partial- and all-in-phase modes, which were unstable, as shown in Fig.~\ref{fig:y_out}(c-2).
As a result, the simulation results indicated that the slope of $l$ in stable synchronization mode was the smallest of the obtained synchronization modes for all $n$.
It is noted that, for $n=3$, the slope of $l$ in the partial-in-phase mode was almost equal to that in the three-phase mode [Fig.~\ref{fig:y_out}(b-2)], whereas the oscillatory amplitudes in these modes were significantly different [Fig.~\ref{fig:y_out}(b-1)].
Thus, another criterion that is related to the oscillatory amplitude could be considered for the stable synchronization modes, besides the small absolute changes in $y_\mathrm{out}$.

\section{conclusion}

We performed two-dimensional hydrodynamic simulation for $n$ coupled identical density oscillators ($n=1,2,3,4$).
The antiphase ($n=2$), three-phase ($n=3$), and 2-2 partial-in-phase ($n=4$) synchronization modes appeared, which agreed with the experimental observation reported in previous studies.
The all-in-phase and partial-in-phase modes were also realized by setting the identical initial conditions.
The stabilities of the synchronization modes were analyzed based on the phase reduction theory, where the phase response to the perturbation was obtained from the simulation for a single density oscillator.
The antiphase ($n=2$), three-phase ($n=3$), and 2-2 partial-in-phase ($n=4$) modes were found to be stable, whereas the other modes were found to be unstable.
The linear stability analysis with the Fourier components of the phase coupling function well reproduced these stabilities.
We numerically confirmed that the stable synchronization modes indicated smaller absolute changes in the water levels in the outer containers than unstable modes, which could be one of the criteria for selection of the synchronization mode in the coupled density oscillators.

Our simulation for coupled density oscillators is suitable to investigate various synchronization phenomena by changing the parameter or boundary condition.
We expect that studies on the hydrodynamic behavior and phase dynamics of coupled density oscillators, including the present study, will contribute to further understanding of the synchronization phenomena in fluid systems.

\section*{acknowledgments}

This work was supported by JST SPRING, Grant No. JPMJSP2109 (N.T.), by JSPS KAKENHI Grants No. JP19H00749, No. JP21K13891 (H.I.), No. JP20H02712, No. JP21H00996, and No. JP21H01004 (H.K.), and by the Cooperative Research Program of ``NJRC Mater. \& Dev.'' No. 20224003 (H.K.).
This work was also supported by JSPS and MESS Japan-Slovenia Research Cooperative Program Grant No. JPJSBP120215001 (H.I.), and JSPS and PAN under the Japan-Poland Research Cooperative Program No. JPJSBP120204602 (H.K.).

\section*{APPENDIX A: FOURIER COMPONENTS OF $Z(\phi)$ AND $f(\phi)$}
\setcounter{equation}{0}
\renewcommand{\theequation}{A\arabic{equation}}

\begin{figure}[tb]
  \begin{center}
    \includegraphics{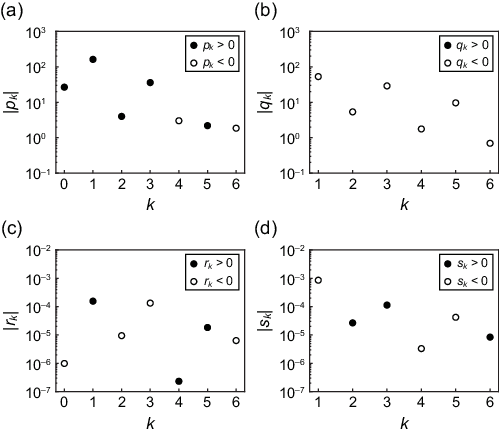}
  \end{center}
  \caption{Fourier components of $Z(\phi)$ and $f(\phi)$. (a) Fourier cosine coefficient $p_k$ of $Z(\phi)$. (b) Fourier sine coefficient $q_k$ of $Z(\phi)$. (c) Fourier cosine coefficient $r_k$ of $f(\phi)$. (d) Fourier sine coefficient $s_k$ of $f(\phi)$.}
  \label{fig:appendix}
\end{figure}

We examine the dominant Fourier components of $Z(\phi)$ and $f(\phi)$ and how they relate to $\Gamma(\phi)$.
We consider the Fourier series
\begin{align}
  Z(\phi) &= \frac{p_0}{2} + \sum_{k=1}^{\infty}(p_k \cos(k\phi) + q_k \sin(k\phi)), \\
	f(\phi) &= \frac{r_0}{2} + \sum_{k=1}^{\infty}(r_k \cos(k\phi) + s_k \sin(k\phi)),
\end{align}
and compare the magnitudes of Fourier components for $Z(\phi)$ and $f(\phi)$ in Fig.~\ref{fig:appendix}, which are calculated from the discrete points in Fig.~\ref{fig:response}(c) and \ref{fig:response}(d), respectively.
It was found that the third ($k=3$) Fourier components are the second largest.
We derive the Fourier cosine and sine coefficients of $\Gamma(\phi)$, $a_k$ and $b_k$, respectively, using those of $Z(\phi)$ and $f(\phi)$ as $a_0 = p_0 r_0/2$, $a_k = (p_k r_k + q_k s_k)/2$ ($k\geq 1$), and $b_k = (p_k s_k - q_k r_k)/2$ ($k\geq 1$) based on Eq.~(\ref{eq:gamma}).
The amplitudes of the $k$-th harmonic oscillations for $k\geq 1$, $Z_k$, $f_k$, and $\Gamma_k$, for $Z(\phi)$, $f(\phi)$, and $\Gamma(\phi)$, respectively, are given as
\begin{align}
  Z_k &= \sqrt{p_k^2 + q_k^2}, \\
	f_k &= \sqrt{r_k^2 + s_k^2}, \\
	\Gamma_k &= \sqrt{a_k^2 + b_k^2} \nonumber \\
  &= \frac{1}{2}\sqrt{(p_k r_k + q_k s_k)^2 + (p_k s_k - q_k r_k)^2} \nonumber \\
  &= \frac{Z_k f_k}{2}. \label{eq:product}
\end{align}
The product of $Z_k$ and $f_k$ gives the scale of $\Gamma_k$ by Eq.~(\ref{eq:product}).
The ratio of $\Gamma_1$ and $\Gamma_3$ estimated by the Fourier components of $Z(\phi)$ and $f(\phi)$ in Fig.~\ref{fig:appendix} is $\Gamma_3/\Gamma_1 = (Z_3/ Z_1)(f_3/ f_1) \sim 10^{-1}\times 10^{-1} = 10^{-2}$, which is so small that $\Gamma(\phi)$ looks like a sinusoidal function.
Therefore, we obtained nearly sinusoidal $\Gamma(\phi)$ from nonsinusoidal $Z(\phi)$ and $f(\phi)$.

\section*{APPENDIX B: EIGENVALUES FOR $n=4$ IN A GENERAL CASE}
\setcounter{equation}{0}
\renewcommand{\theequation}{B\arabic{equation}}

We calculate the eigenvalues $\lambda$ of the Jacobian matrix $A$ in Eq.~(\ref{eq:appendix}) for the four-phase and 2-2 partial-in-phase modes ($n=4$) with all Fourier components $a_k$ and $b_k$ of $\Gamma(\phi)$.
The eigenvalues for the four-phase mode are derived as
\begin{align}
  \lambda&=2\left( \sum_{k=1}^\infty A_k b_k \pm i \sum_{k=1}^\infty B_k a_k \right), \, 8\sum_{k=1}^\infty C_k b_k,
\end{align}
where
\begin{align}
  A_k&=
  \begin{cases}
    k \quad &(k=2m-1)\\
    0 \quad &(k=4m-2)\\
    -2k \quad &(k=4m),
  \end{cases}
  \quad m=1,2,\ldots, \\
  B_k&=
  \begin{cases}
    k \quad &(k=4m-3)\\
    -k \quad &(k=4m-1)\\
    0 \quad &(k=2m),
  \end{cases}
  \quad m=1,2,\ldots, \\
  C_k&=
  \begin{cases}
    0 \quad &(k=2m-1)\\
    k/2 \quad &(k=4m-2)\\
    -k/2 \quad &(k=4m),
  \end{cases}
  \quad m=1,2,\ldots.
\end{align}
The eigenvalues for the 2-2 partial-in-phase mode are derived as
\begin{align}
  \lambda&=4\sum_{k=1}^\infty (-1)^{k+1}k b_k, \, -8\sum_{k=1}^\infty D_k b_k \, \text{(duplicate)},
\end{align}
where
\begin{align}
  D_k&=
  \begin{cases}
    0 \quad &(k\neq 2m)\\
    k/2 \quad &(k=2m),
  \end{cases}
  \quad m=1,2,\ldots.
\end{align}
With the numerical result of $b_k$ for $k \leq 6$ shown in Fig.~\ref{fig:components}, the real parts of the eigenvalues are calculated as $\mathrm{Re}(\lambda)=-1.3\times 10^{-1},1.2\times 10^{-4}$ for the four-phase mode and $\mathrm{Re}(\lambda)=-2.7\times 10^{-1},-2.2\times 10^{-4}$ for the 2-2 partial-in-phase mode.
Therefore, the 2-2 partial-in-phase mode is stable, whereas the four-phase mode is unstable.

\end{document}